\documentclass[journal]{IEEEtran}
\usepackage{graphicx}

\usepackage{cite}

\ifCLASSINFOpdf
\else
\fi
\usepackage{amsmath}
\usepackage{amsfonts}
\begin{document}
%
\title{Reversible Embedding to Covers Full of Boundaries}
%
%
%

\author{Hanzhou Wu$^1$, Wei Wang$^1$, Jing Dong$^1$, Yanli Chen$^2$ and Hongxia Wang$^2$\\
$^1$Institute of Automation, Chinese Academy of Sciences, Beijing 100190, China\\
$^2$School of Information Science and Technology, Southwest Jiaotong University, Chengdu 611756, China
}

%
%

\markboth{}%
{}
%



\maketitle

\begin{abstract}
In reversible data embedding, to avoid overflow and underflow problem, before data embedding, boundary pixels are recorded as side information, which may be losslessly compressed. The existing algorithms often assume that a natural image has little boundary pixels so that the size of side information is small. Accordingly, a relatively high pure payload could be achieved. However, there actually may exist a lot of boundary pixels in a natural image, implying that, the size of side information could be very large. Therefore, when to directly use the existing algorithms, the pure embedding capacity may be not sufficient. In order to address this problem, in this paper, we present a new and efficient framework to reversible data embedding in images that have lots of boundary pixels. The core idea is to losslessly preprocess boundary pixels so that it can significantly reduce the side information. Experimental results have shown the superiority and applicability of our work.
\end{abstract}


%
\IEEEpeerreviewmaketitle

\section{Motiviation}
\IEEEPARstart{R}{eversible} data embedding, also called reversible data hiding \cite{tian:de}, is a special fragile technique that could benefit sensitive applications that require no distortion of the cover. It works by hiding a message such as authentication data within a cover by slightly altering the cover. At the decoder side, one can extract the hidden data from the marked content. And, the original content can be perfectly reconstructed.

The reversible data embedding can be modeled as a rate-distortion optimization problem. We hope to embed as many message bits as possible while the introduced distortion is low. A number of algorithms have been proposed in the past years. Early algorithms use lossless compression to vacate room for data embedding. More efficient approaches are introduced to increase the data embedding capacity or reduce the distortion such as difference expansion \cite{tian:de}, histogram shifting \cite{ni:hs} and other methods \cite{ma:cdm, hzwu:iwdw}. Nowadays, advanced algorithms use prediction-errors (PEs) \cite{hsu:smp, ou:pairwise, li:mhs, hzwu:DCSPF, hzwu:PPE} of the cover to hide secret data since PEs could provide superior rate-distortion performance.

In order to avoid underflow and overflow problem during data embedding, boundary pixels should be adjusted into the reliable range and recorded as side information, which will be embedded into the cover image together with the secret data. The existing algorithms often assume that, the used cover image is natural and thus the size of side information could be small, which will have little impact on the pure embedding capacity. However, even for natural images, there may exist a lot of boundary pixels. This implies that, the side information may have significant impact on the pure embedding capacity.

We use the image database BOSSBase \cite{bas:bossbase} for explanation. We assume that the pixels with a value of 0/255 are boundary pixels. Fig. 1 shows the number of boundary pixels in each image. It is observed that, there are many images that have lots of boundary pixels, implying that the corresponding side information may require a lot of bits. In reversible data embedding, a commonly used operation to construct the side information is first to assign one bit to each pixel representing whether the present pixel is a boundary or not. Then, the resulting binary matrix also called the location map is losslessly compressed. This operation is effective when the number of boundary pixels is small. However, it may have poor performance in images full of boundary pixels, especially for the case that boundary pixels are widely distributed. Fig. 2 shows an example. Regardless of the lossless compression algorithm, the compression ratio for the location map would be intuitively very low.

\begin{figure}[!t]
\centering
\includegraphics[width=3.4in]{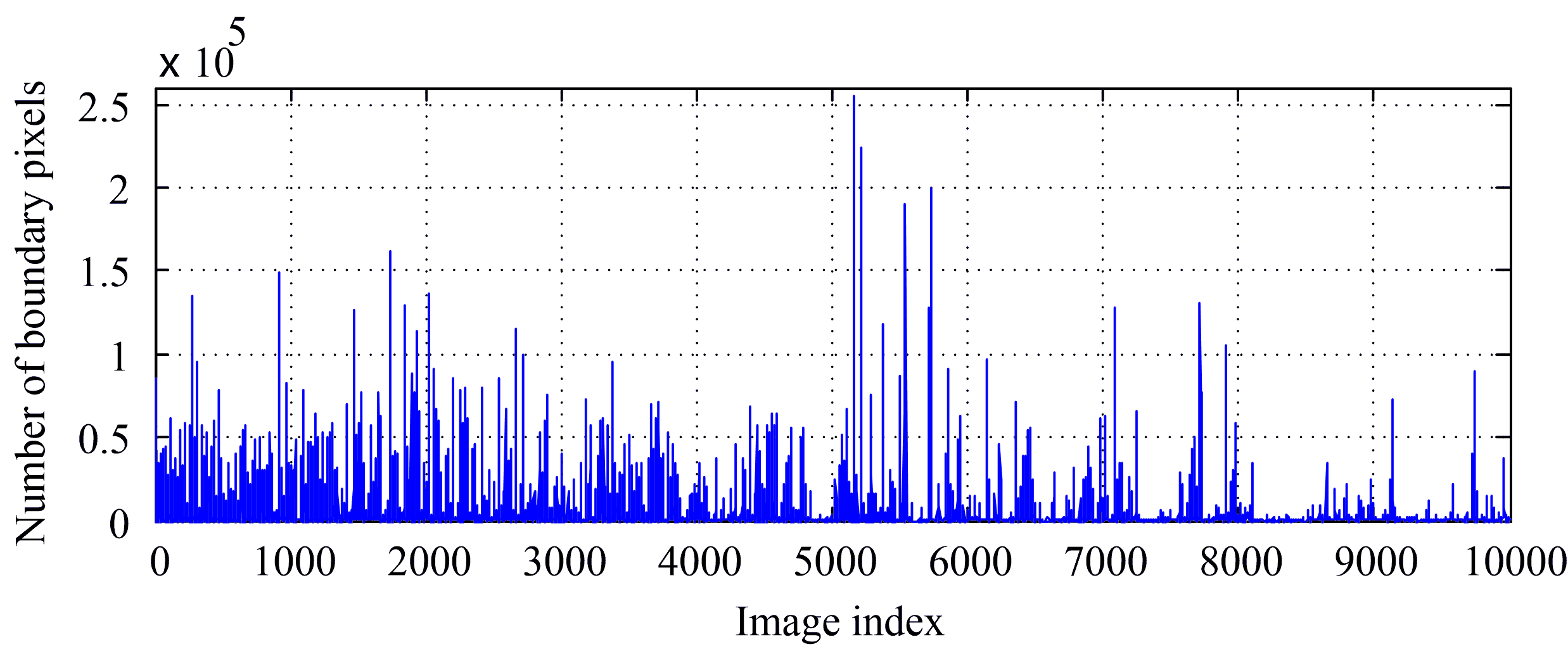}
\caption{The number of boundary pixels in each image.}
\end{figure}

\begin{figure}[!t]
\centering
\includegraphics[width=3.2in]{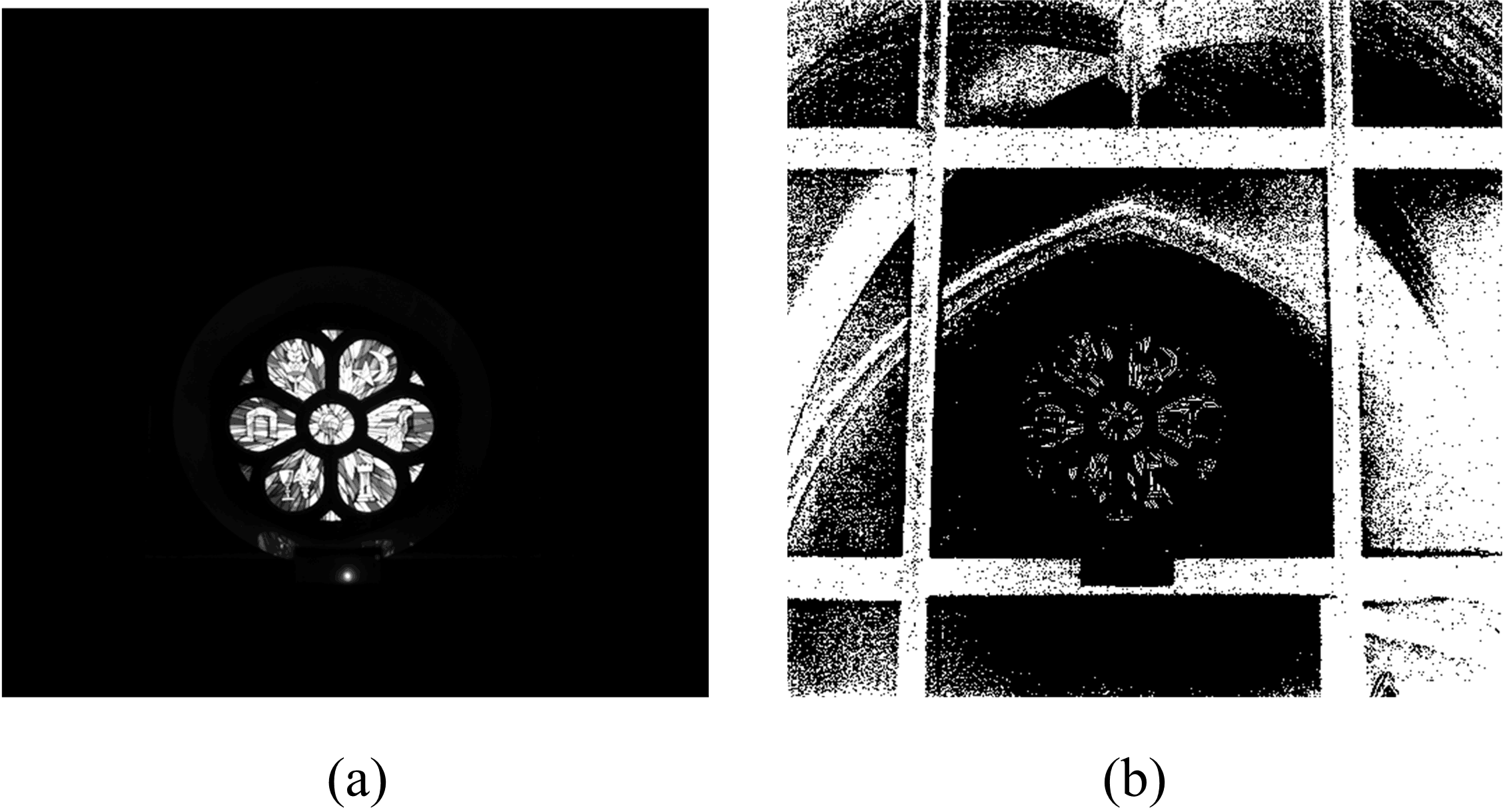}
\caption{An image full of boundary pixels selected from BOSSBase. (a) 5727.pgm, (b) the location map where the boundary pixels are in white area.}
\end{figure}

We sort all images in BOSSBase according to the number of the boundary pixels. We choose the top-200 largest number of boundary pixels of images out and construct the losslessly compressed location maps by arithmetic coding. Fig. 3 shows the size of the losslessly compressed location map for each image. It is observed that, the sizes of the compressed location maps are all very large, indicating that, many existing algorithms may carry a very low pure payload and even cannot carry extra bits for those natural images full of boundary pixels. This has motivated the authors in this paper to propose an efficient algorithm to address this very important problem.

The rest are organized as follows. In Section II, we introduce the proposed reversible data embedding framework for images that have lots of boundary pixels. Then, we conduct experiments to show the performance in Section III. Finally, we conclude this paper in Section IV.

\begin{figure}[!t]
\centering
\includegraphics[width=3.4in]{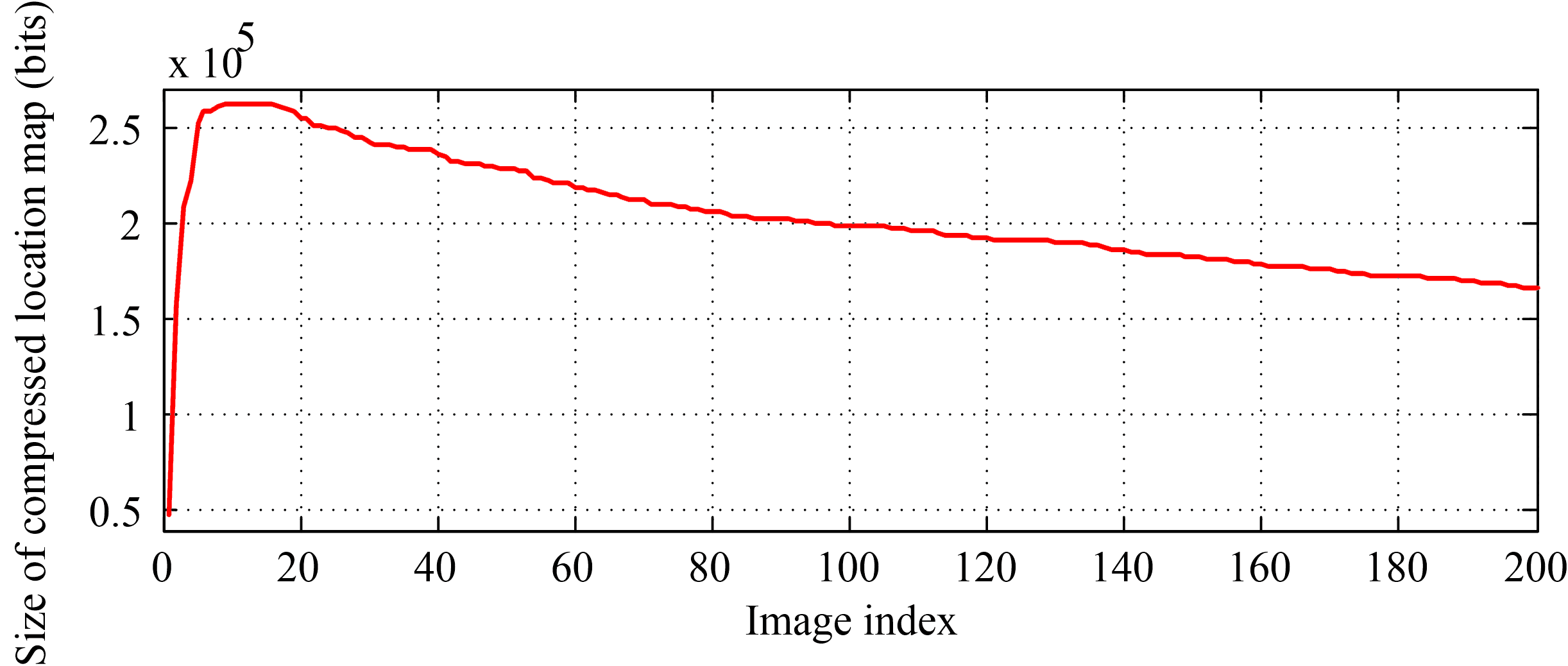}
\caption{The size of losslessly compressed location map for each of the selected images. Index 1 means the image has the largest number of boundary pixels.}
\end{figure}

\section{Proposed Framework}
The proposed work involves three steps. First, all pixels are preprocessed by prediction, for which the number of boundary pixels will be significantly reduced, resulting in a small size of the compressed location map. Then, any suitable reversible embedding operation can be applied to the preprocessed image to carry a payload. Finally, the data extraction and image recovery can be performed similarly.

\subsection{Prediction-based Preprocessing}
Let $\mathcal{O} = \mathcal{I}^{n\times m}$ denote the original image sized $n\times m$ with the pixel range $\mathcal{I} = \{0, 1, ..., 255\}$. For compactness, we sometimes consider $\mathcal{O}$ as the set including all pixels and say ``pixel $o_{i,j}$'' meaning a pixel located at position $(i, j)$ whose value is $o_{i,j}$. $o_{i,j}\in \mathcal{O}$ is called as a boundary pixel if
\begin{equation*}
o_{i,j}\in [0, T)\cup(255-T, 255],
\end{equation*}
where $T> 0$ is a predetermined parameter relying on the data embedding operation. It is always assumed that $T << 255 - T$.

We are to preprocess $\mathcal{O}$ to generate a new image $\mathcal{X}$ and a location map $\mathcal{L}$. First, we divide $\mathcal{O}$ into two subsets, i.e.,
\begin{equation*}
\mathcal{O}_b = \{o_{i,j}\in\mathcal{O}~|~(i+j)~\text{mod}~2~=~b\}, b \in \{0,1\}.
\end{equation*}

Then, we use $\mathcal{O}_1$ to predict $\mathcal{O}_0$. In detail, for each $o_{i,j}\in \mathcal{O}_0$, we determine its prediction value by:
\begin{equation}
\hat{o}_{i,j} = \left [\frac{o_{i-1,j}+o_{i+1,j}+o_{i,j-1}+o_{i,j+1}}{4} \right ].
\end{equation}
where $[\cdot]$ returns a nearest integer. It is easy to modify Eq. (1) slightly in case that a pixel position is out of the image.

We use a threshold $t_0$ to generate an image $\mathcal{X}^{(0)}$, i.e.,
\begin{equation}
\begin{split}
x_{i,j}^{(0)} = \left\{\begin{matrix}
o_{i,j} + T,~&\text{if}~\hat{o}_{i,j}<t_0~\text{and}~o_{i,j}\in \mathcal{O}_0,\\
o_{i,j} - T,~&\text{if}~\hat{o}_{i,j}>255-t_0~\text{and}~o_{i,j}\in \mathcal{O}_0,\\
o_{i,j},~&\text{otherwise}.
\end{matrix}\right.
\end{split}
\end{equation}

The principle behind Eq. (2) is that, if the prediction value of a pixel is close to the boundary value, its original value should be close to the boundary value as well. Fig. 4 shows two examples by using the predictor in Eq. (1). It is seen that, for both images, there exist strong correlations between the original values and the prediction values. Thus, we can adjust the raw value into the reliable range according to its prediction value. For each $o_{i,j}\in \mathcal{O}_1$, we continue to compute its prediction value in $\mathcal{X}^{(0)}$ by:
\begin{equation}
\hat{x}_{i,j}^{(0)} = \left [\frac{x_{i-1,j}^{(0)}+x_{i+1,j}^{(0)}+x_{i,j-1}^{(0)}+x_{i,j+1}^{(0)}}{4} \right ].
\end{equation}

We use a threshold $t_1$ to generate another image $\mathcal{X}^{(1)}$ from $\mathcal{X}^{(0)}$ by:
\begin{equation}
\begin{split}
x_{i,j}^{(1)} = \left\{\begin{matrix}
x_{i,j}^{(0)} + T,&\text{if}~\hat{x}_{i,j}^{(0)}<t_1~\text{and}~o_{i,j}\in \mathcal{O}_1,\\
x_{i,j}^{(0)} - T,&\text{if}~\hat{x}_{i,j}^{(0)}>255-t_1~\text{and}~o_{i,j}\in \mathcal{O}_1,\\
x_{i,j}^{(0)},&\text{otherwise}.
\end{matrix}\right.
\end{split}
\end{equation}

The pixel values of $\mathcal{X}^{(1)}$ must be in range $[-T, 255 + T]$. We will adjust the pixels in $\mathcal{X}^{(1)}$ into the range $[T, 255-T]$ to generate the final image $\mathcal{X}$. In detail, for all possible $x_{i,j}\in\mathcal{X}$, we compute it as follows:
\begin{equation}
\begin{split}
x_{i,j} = \left\{\begin{matrix}
T,~&\text{if}~x_{i,j}^{(1)}<T,\\
255 - T,~&\text{if}~x_{i,j}^{(1)}>255-T,\\
x_{i,j}^{(1)},~&\text{otherwise}.
\end{matrix}\right.
\end{split}
\end{equation}

$\mathcal{X}$ will not contain boundary pixels. We record such pixel positions $(i,j)$ that $x_{i,j}^{(1)}\in [-T, T)\cup (255-T, 255+T]$. We construct a $(2T+1)$-ary location map $\mathcal{L}=\{0, 1, ..., 2T\}^{n\times m}$ to address this issue, i.e.,
\begin{equation}
\begin{split}
l_{i,j} = \left\{\begin{matrix}
x_{i,j}^{(1)}+T,~&\text{if}~x_{i,j}^{(1)}<T,\\
255 + T - x_{i,j}^{(1)},~&\text{if}~x_{i,j}^{(1)}>255-T,\\
2T,~&\text{otherwise}.
\end{matrix}\right.
\end{split}
\end{equation}
where $l_{i,j}\in \mathcal{L}$ is corresponding to the pixel $x_{i,j}\in \mathcal{X}$.

\begin{figure}[!t]
\centering
\includegraphics[width=3.5in]{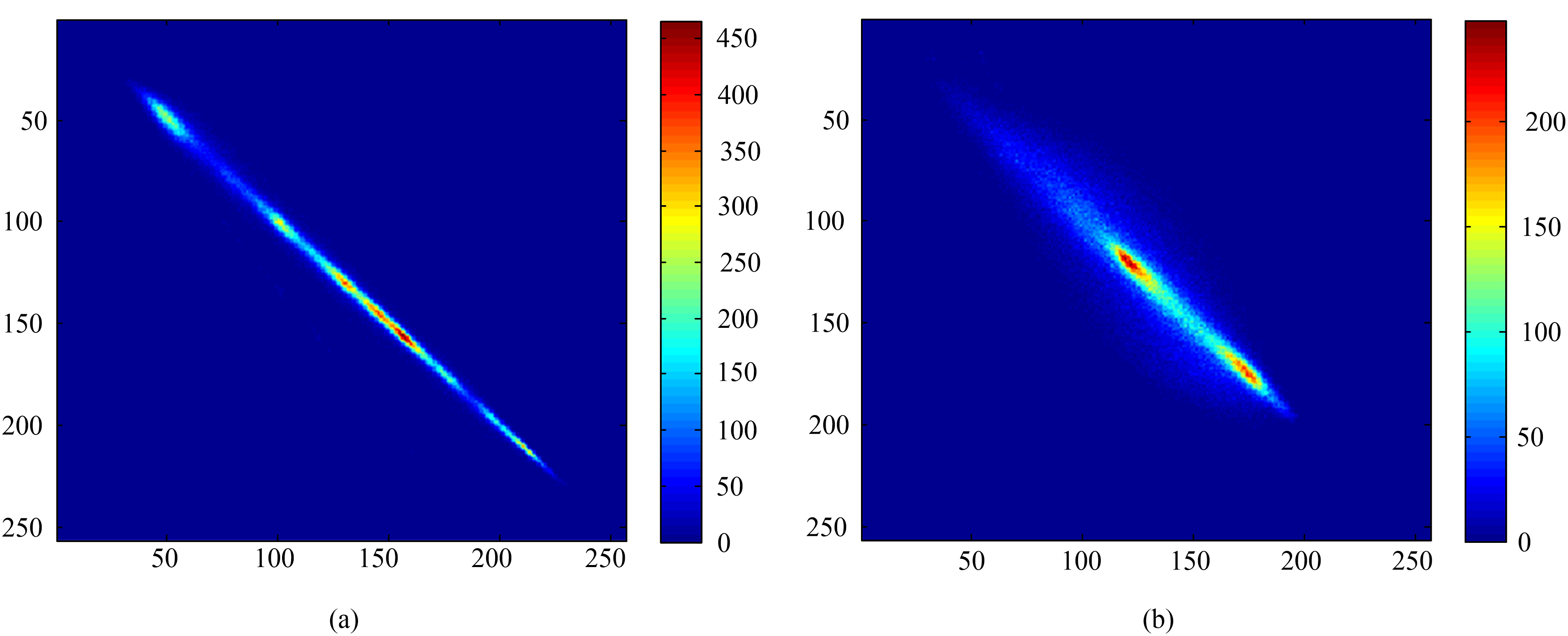}
\caption{The statistical distribution between the pixel values and their prediction values. (a) Lena, (b) Baboon. Both are grayscale and sized $512\times 512$.}
\end{figure}

\subsection{Reversible Data Embedding}
We embed the required payload $\mathcal{P}$ and the losslessly compressed $\mathcal{L}$ into $\mathcal{X}$, rather than $\mathcal{O}$. It is seen that, there has no need to construct a new location map since $\mathcal{X}$ does not contain boundary pixels. It is inferred that, one can use many existing state-of-the-art algorithms for reversible data embedding since the data embedding operation here is open to design. We will not focus on the detailed reversible embedding operation.

\subsection{Data Extraction and Image Recovery}
Suppose that, we have embedded $\mathcal{P}$, $\mathcal{L}$ and data embedding parameters into $\mathcal{X}$, resulting in a marked image $\mathcal{Y}$. Notice that, $\mathcal{L}$ has been compressed in advance. For a receiver, he needs to extract the embedded data and reconstruct $\mathcal{O}$ from $\mathcal{Y}$. It is straightforward to reconstruct $\mathcal{P}$, $\mathcal{L}$ and $\mathcal{X}$ from $\mathcal{Y}$. Our goal is to reconstruct $\mathcal{O}$. With Eqs. (5, 6), it is straightforward to reconstruct $\mathcal{X}^{(1)}$ from $\mathcal{X}$ and $\mathcal{L}$ by:
\begin{equation}
\begin{split}
x_{i,j}^{(1)} = \left\{\begin{matrix}
x_{i,j},&\text{if}~l_{i,j}=2T,\\
l_{i,j}-T,&\text{if}~l_{i,j}\neq 2T~\text{and}~x_{i,j}=T,\\
255 + T - l_{i,j},&\text{otherwise}.
\end{matrix}\right.
\end{split}
\end{equation}

Thereafter, we reconstruct $\mathcal{X}^{(0)}$ from $\mathcal{X}^{(1)}$. First, we initialize $\mathcal{X}^{(0)} = \mathcal{X}^{(1)}$. Then, we predict all $x_{i,j}^{(0)}$ corresponding to $\mathcal{O}_1$ by Eq. (3). According to $t_1$, $\mathcal{X}^{(0)}$ is finally identified by:
\begin{equation}
\begin{split}
x_{i,j}^{(0)} = \left\{\begin{matrix}
x_{i,j}^{(1)} - T,&\text{if}~\hat{x}_{i,j}^{(0)}<t_1~\text{and}~o_{i,j}\in \mathcal{O}_1,\\
x_{i,j}^{(1)} + T,&\text{if}~\hat{x}_{i,j}^{(0)}>255-t_1~\text{and}~o_{i,j}\in \mathcal{O}_1,\\
x_{i,j}^{(1)},&\text{otherwise}.
\end{matrix}\right.
\end{split}
\end{equation}

Similarly, we initialize $\mathcal{O} = \mathcal{X}^{(0)}$. We predict all $o_{i,j}\in \mathcal{O}_0$ by Eq. (1). With $t_0$, $\mathcal{O}$ can be reconstructed as:
\begin{equation}
\begin{split}
o_{i,j} = \left\{\begin{matrix}
x_{i,j}^{(0)} - T,~&\text{if}~\hat{o}_{i,j}<t_0~\text{and}~o_{i,j}\in \mathcal{O}_0,\\
x_{i,j}^{(0)} + T,~&\text{if}~\hat{o}_{i,j}>255-t_0~\text{and}~o_{i,j}\in \mathcal{O}_0,\\
x_{i,j}^{(0)},~&\text{otherwise}.
\end{matrix}\right.
\end{split}
\end{equation}

Therefore, $\mathcal{P}$ and $\mathcal{O}$ can be perfectly reconstructed. Notice that, $t_0$, $t_1$ and $T$ are parameters that should be embedded into $\mathcal{X}$ previously.
\begin{figure}[!t]
\centering
\includegraphics[width=3.4in]{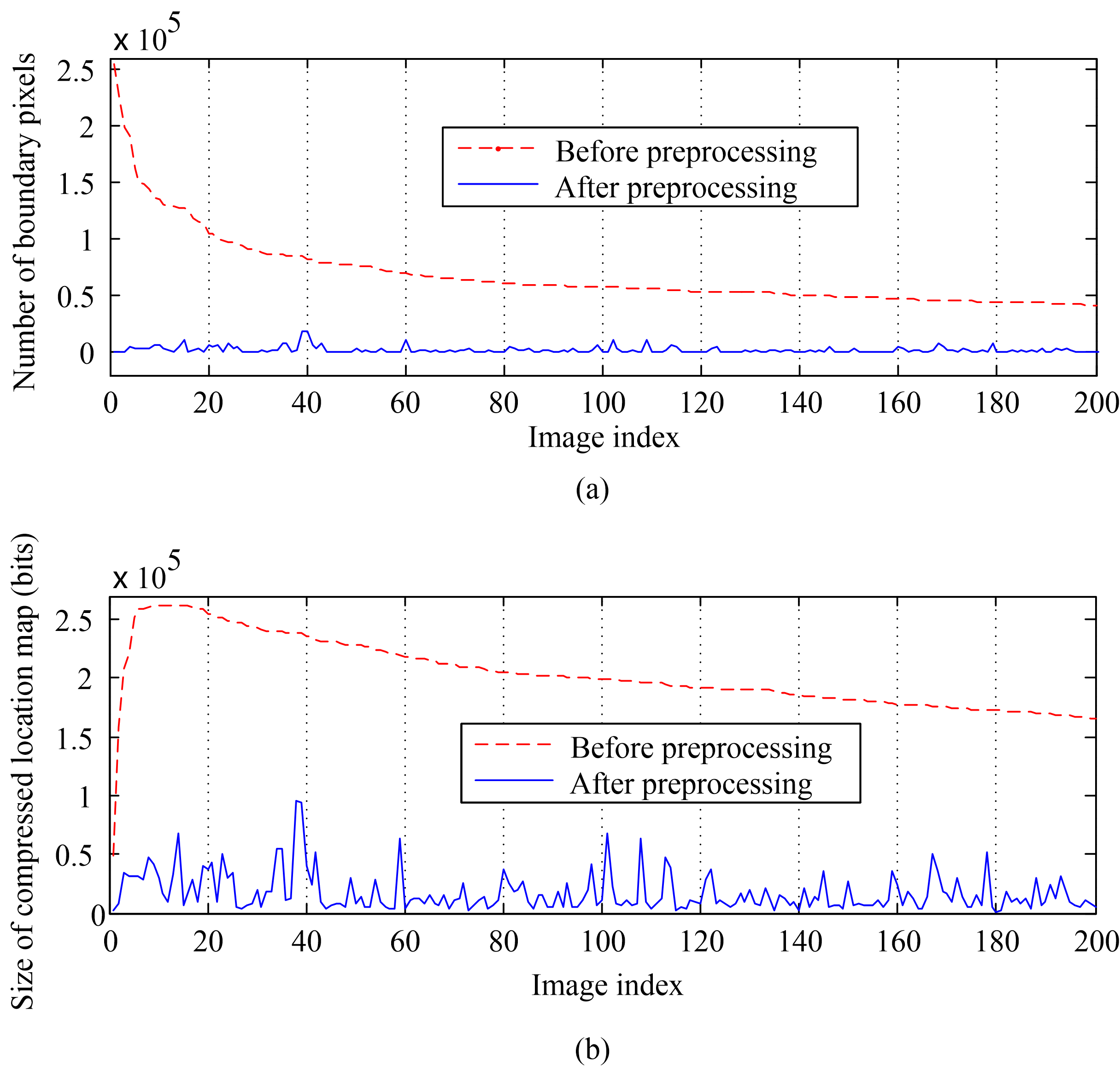}
\caption{Comparison for the number of boundary pixels as well as the size of compressed location map before/after preprocessing: $t_0 = 1$ and $t_1 = 4$.}
\end{figure}

\section{Performance Evaluation and Analysis}
The core contribution of our work is that, we introduce an efficient losslessly processing technique to significantly reduce the sizes of location maps for images full of boundary pixels.
To verify the performance, we choose the 200 images mentioned in Fig. 3 for experiments. For an original image $\mathcal{O}$, we set $T = 1$ and use arithmetic coding to losslessly compress the corresponding location map. For the corresponding $\mathcal{X}$, we define $x_{i,j}\in \mathcal{X}$ as a boundary pixel if $l_{i,j}\neq 2T$. We first use $t_0 = 1$ and $t_1 = 4$ to compare the compression performance.

As shown in Fig. 5 (a), the number of boundaries is significantly reduced. In Fig. 5, ``before preprocessing'' corresponds to $\mathcal{O}$ and the other one corresponds to $\mathcal{X}$. As shown in Fig. 5 (b), the size of location map is significantly reduced as well, meaning that, a sufficient pure payload could be carried. We define two ratios as follows:
\begin{equation}
r_0 = \frac{\text{Number of boundaries in}~\mathcal{X}}{\text{Number of boundaries in}~\mathcal{O}}\times 100\%,
\end{equation}
and
\begin{equation}
r_1 = \frac{\text{Size of compressed}~\mathcal{L}}{\text{Size of compressed location map for}~\mathcal{O}}\times 100\%.
\end{equation}

We compute the mean value of $r_0$ and $r_1$ for the 200 images. They are $3.40\%$ and $8.21\%$, respectively. It has shown that our work can significantly reduce side information, which would be quite helpful for subsequent data embedding operation.
\begin{figure}[!t]
\centering
\includegraphics[width=3.4in]{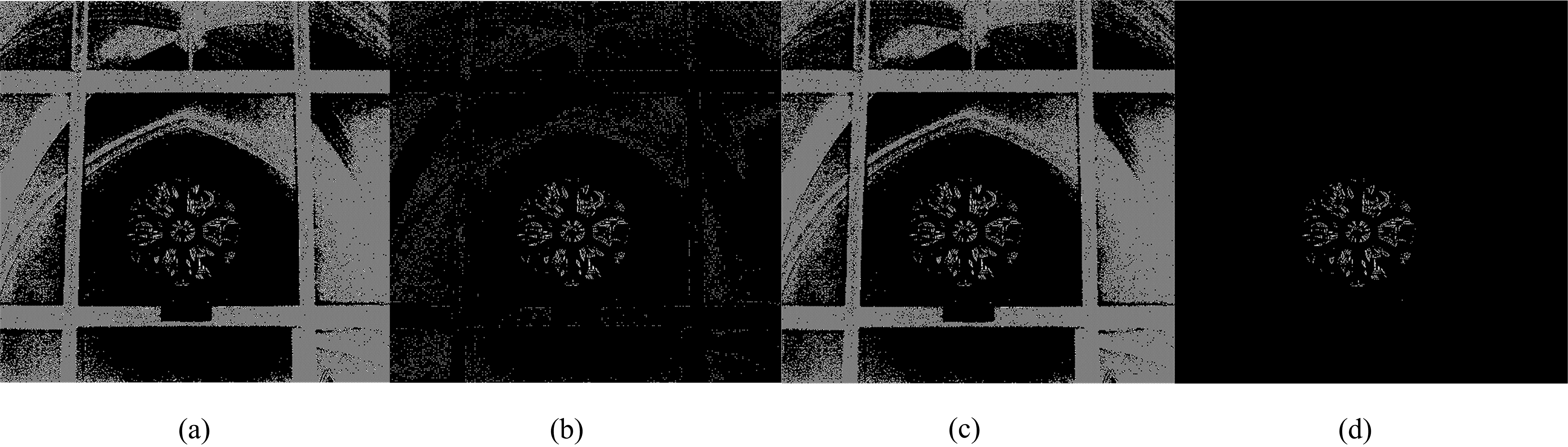}
\caption{Different location maps due to different $t_0$ and $t_1$: (a) $t_0 = t_1 = 1$, (b) $t_0 = 1, t_1 = 4$, (c) $t_0 = 4, t_1 = 1$, and (d) $t_0 = t_1 = 4$. The boundary pixels are in the white area.}
\end{figure}

\begin{table}[h]
        \caption{The mean values of $r_0$ (\%) due to different $t_0$ and $t_1$}
		\centering
		\begin{tabular}{|l|c|c|c|c|c|}\hline
			   & $t_1=1$ & $t_1=2$ & $t_1=4$ & $t_1=8$ & $t_1=16$\\\hline
			$t_0=1$  & 52.83 & 4.26 & 3.40 & 2.84 & 2.30\\
			$t_0=2$  & 52.17 & 3.79 & 2.47 & 1.91 & 1.37\\
			$t_0=4$  & 52.05 & 3.65 & 2.29 & 1.72 & 1.18\\
			$t_0=8$  & 51.90 & 3.50 & 2.13 & 1.55 & 1.00\\
			$t_0=16$ & 51.73 & 3.34 & 1.96 & 1.38 & 0.83\\\hline
		\end{tabular}
\end{table}

\begin{table}[h]
        \caption{The mean values of $r_1$ (\%) due to different $t_0$ and $t_1$}
		\centering
		\begin{tabular}{|l|c|c|c|c|c|}\hline
			   & $t_1=1$ & $t_1=2$ & $t_1=4$ & $t_1=8$ & $t_1=16$\\\hline
			$t_0=1$  & 73.08 & 9.89 & 8.21 & 7.10 & 5.98\\
			$t_0=2$  & 72.58 & 8.93 & 6.14 & 4.95 & 3.75\\
			$t_0=4$  & 72.48 & 8.68 & 5.75 & 4.51 & 3.27\\
			$t_0=8$  & 72.36 & 8.40 & 5.41 & 4.12 & 2.83\\
			$t_0=16$ & 72.22 & 8.08 & 5.07 & 3.74 & 2.39\\\hline
		\end{tabular}
\end{table}

Actually, different $t_0$ and $t_1$ will result in different performance. Fig. 6 shows the different location maps for the image shown in Fig. 2 due to different $t_0$ and $t_1$. It is observed that $t_0 < t_1$ results in a smaller number of boundary pixels. The reason is that, when to predict the pixels corresponding to $\mathcal{O}_1$, the contexts are prediction values resulting in degradation of the prediction accuracy. To further evaluate the impact due to different $t_0$ and $t_1$, we perform experiments on the 200 images. Table I and II shows the mean values of $r_0$ and $r_1$. It has verified our perspective. And, it is suggested to use $t_1 \geq 2$.

With a preprocessed image $\mathcal{X}$, we need to embed $\mathcal{P}$ together with the compressed $\mathcal{L}$ and other parameters. We focus on the data embedding capacity (bits per pixel, bpp):
\begin{equation}
r_\text{emb} = |\mathcal{P}_\text{max}|/|\mathcal{O}|,
\end{equation}
where $|\mathcal{P}_\text{max}|$ represents the size of the maximum embeddable payload $\mathcal{P}_\text{max}$ and $|\mathcal{O}|$ shows the total number of pixels.

One can apply any efficient data embedding algorithms. We use the methods presented in \cite{hsu:smp}, \cite{li:mhs} and \cite{hzwu:PPE} for experiments to evaluate rate-distortion performance. The PSNR is considered as the distortion measure, and is determined between $\mathcal{O}$ and $\mathcal{Y}$. We choose the top-40 largest number of boundary pixels of images for experiments. The image indexes in BOSSBase have been given in Table III. We compare $r_\text{emb}$ and the corresponding distortion for both the original image and the corresponding preprocessed image. We vary $t_0$ and $t_1$ from 1 to 16 by a step of 1 for optimization. The one resulting in the maximum $r_\text{emb}$ will be selected as the result since a data hider always has the freedom to choose $t_0$ and $t_1$. During the data embedding, for the original image, the boundary pixels are recorded by a location map losslessly compressed by arithmetic coding. For fair comparison, the corresponding location maps for the preprocessed images are compressed by arithmetic coding as well. Thereafter, both the original image and the preprocessed image are embedded.

\begin{table}[h]
        \caption{The test image indexes.}
		\centering
		\begin{tabular}{|c|c|c|c|c|c|c|c|}\hline
			  5162 & 5215 & 5726 & 5543 & 1732 & 5542 &  922 & 5723 \\
              2025 &  269 & 7718 & 1851 & 5718 & 5716 & 7082 & 1465 \\
              5370 & 2668 & 1933 & 5727 & 7912 & 2724 & 5722 & 6147 \\
              5717 &  318 & 3382 & 2066 & 5859 & 9740 & 1890 & 5493 \\
              5371 & 5729 & 5724 &    1 & 7091 & 2202 & 2547 & 5541 \\\hline
		\end{tabular}
\end{table}

\begin{table}[h]
        \caption{The number of embeddable images among 40 test images.}
		\centering
		\begin{tabular}{|c|c|c|c|c|c|c|c|c|c|c|c|}\hline
              \multicolumn{2}{|c|}{Method in \cite{hsu:smp}}&\multicolumn{2}{c|}{Method in \cite{li:mhs}}&\multicolumn{2}{c|}{Method in \cite{hzwu:PPE}}\\\hline
			  before & after & before & after & before & after\\\hline
              5 & 40 & 5 & 40 & 5 & 40\\\hline
		\end{tabular}
\end{table}

\begin{table}[h]
        \caption{$r_\text{emb}$ (bpp) for the embeddable images.}
		\centering
		\begin{tabular}{|c|c|c|c|c|c|c|c|c|c|c|c|c|}\hline
              / &\multicolumn{2}{|c|}{Method in \cite{hsu:smp}}&\multicolumn{2}{c|}{Method in \cite{li:mhs}}&\multicolumn{2}{c|}{Method in \cite{hzwu:PPE}}\\\hline
			  image & before & after & before & after & before & after\\\hline
              5162.pgm & 0.79 & 0.98 & 0.79 & 0.97 & 0.78 & 0.96\\
              5215.pgm & 0.38 & 0.97 & 0.37 & 0.97 & 0.36 & 0.95\\
              5726.pgm & 0.15 & 0.86 & 0.14 & 0.88 & 0.13 & 0.85\\
              5543.pgm & 0.04 & 0.84 & 0.06 & 0.88 & 0.04 & 0.84\\
              5541.pgm & 0.02 & 0.87 & 0.04 & 0.90 & 0.02 & 0.87\\\hline
		\end{tabular}
\end{table}

Experimental results show that, there exist a large ratio of images that the existing algorithms cannot be applied to them directly, namely, $r_\text{emb} = 0$ for the original images since the size of compressed location maps are too large to embed extra bits. As shown in Table IV, we count the number of embeddable images (i.e., $r_\text{emb} > 0$). In Table IV, ``before'' corresponds to the original image, and ``after'' corresponds to the preprocessed image. It can be observed that, the proposed work significantly improves the ability of carrying additional data for the existing works. We compare $r_\text{emb}$ and PSNR for those embeddable images. Table V and VI show the results. It can be seen that, the proposed work significantly increases the capacity, and provides a high image quality. We further determine the mean value of $r_\text{emb}$ and PSNR for the 40 test images using the three data embedding algorithms equipped with the preprocessing technique. Table VII provides the results, which has implied that, the proposed work has good ability to improve the rate-distortion performance of many existing algorithms on images containing lots of boundary pixels.

\begin{table}[h]
        \caption{The PSNRs (dB) for the marked images.}
		\centering
		\begin{tabular}{|c|c|c|c|c|c|c|c|c|c|c|c|c|}\hline
              / &\multicolumn{2}{|c|}{Method in \cite{hsu:smp}}&\multicolumn{2}{c|}{Method in \cite{li:mhs}}&\multicolumn{2}{c|}{Method in \cite{hzwu:PPE}}\\\hline
			  image & before & after & before & after & before & after\\\hline
              5162.pgm & 44.23 & 44.23 & 44.35 & 46.61 & 45.61 & 45.60\\
              5215.pgm & 44.68 & 44.25 & 47.00 & 46.58 & 46.00 & 45.58\\
              5726.pgm & 45.05 & 44.41 & 47.27 & 46.69 & 46.31 & 45.56\\
              5543.pgm & 45.17 & 44.52 & 47.30 & 46.73 & 46.38 & 45.60\\
              5541.pgm & 47.49 & 44.49 & 49.07 & 46.63 & 48.35 & 45.44\\\hline
		\end{tabular}
\end{table}

\begin{table}[h]
        \caption{The mean value of $r_\text{emb}$ (bpp) and PSNR (dB) for the 40 test images using the proposed preprocessing technique.}
		\centering
		\begin{tabular}{|c|c|c|c|c|c|c|c|c|c|c|c|}\hline
              \multicolumn{2}{|c|}{Method in \cite{hsu:smp}}&\multicolumn{2}{c|}{Method in \cite{li:mhs}}&\multicolumn{2}{c|}{Method in \cite{hzwu:PPE}}\\\hline
			  $r_\text{emb}$ & PSNR & $r_\text{emb}$ & PSNR & $r_\text{emb}$ & PSNR\\\hline
              0.63 & 46.69 & 0.67 & 47.16 & 0.63 & 46.75\\\hline
		\end{tabular}
\end{table}

\section{Conclusion and Discussion}
In practice, it is quite easy to acquire images full of boundary pixels such as medical images, remote sensing images and natural sceneries, e.g., white cloud and dark night. The existing works often focus on images having little boundary pixels and provide superior rate-distortion performance on them. However, they may not work very well for images full of boundaries. In this paper, we present an efficient losslessly preprocessing algorithm to reversible data embedding in images that contain lots of boundary pixels. The reversible embedding operation for the proposed work is open to design. Experimental results have shown that our work could significantly reduce the size of the side information, which can benefit reversible data embedding performance a lot. The proposed work could have good potential in reversible data embedding. The future work is to design data embedding algorithms that can well exploit the statistical characteristics of the preprocessed image.

\ifCLASSOPTIONcaptionsoff
  \newpage
\fi

\end{document}